\documentclass[prd,twocolumn,floats,floatfix,showpacs,nofootinbib]{revtex4}
\usepackage{graphicx}
\usepackage{dcolumn}
\usepackage{bm}
\usepackage{graphics}
\usepackage{slashed}
\usepackage{amssymb}
\usepackage{natbib}
\usepackage{amsmath}
\usepackage{url}
\usepackage{color}
\newcommand{\bea}{\begin{eqnarray}}
\newcommand{\ena}{\end{eqnarray}}
\newcommand{\bean}{\begin{eqnarray*}}
\newcommand{\enan}{\end{eqnarray*}}

\begin{document}

\title{The gravitational mechanism to generate mass II}

\author{M. Novello \footnote{M. Novello is Cesare Lattes ICRANet Professor}} \email{novello@cbpf.br}
\author{E. Bittencourt}\email{eduhsb@cbpf.br}
\affiliation{Instituto de Cosmologia Relatividade Astrofisica ICRA -
CBPF\\ Rua Dr. Xavier Sigaud, 150, CEP 22290-180, Rio de Janeiro,
Brazil}
\pacs{98.80.Cq}
\date{\today}

\begin{abstract}
With the eminent confirmation or disproof of the existence of Higgs
boson by experiments on the LHC it is time to analyze in a
non-dogmatic way the suggestions to understand the origin of the mass.
Here we analyze the recent proposal according to which gravity is
what is really responsible for the generation of mass of all bodies.
The great novelty of such mechanism is that the gravitational field
acts merely as a catalyst, once the final expression of the mass
does not depend either on the intensity or on the particular characteristics of
the gravitational field.
\end{abstract}

\maketitle

\section{Introduction: recipe for generating mass}

\vspace{0.50cm}

In order to become a  reliable candidate as a mechanism to generate
mass, there are three indispensable conditions that such mechanism
has to fulfil, to wit:

\begin{itemize}
\item{There must exist a universal field that interacts with
all kinds of particles;}
\item{There must exist a free parameter such that different bodies
can acquire distinct values for their corresponding mass (the
spectrum of mass);}
\item{This field must be such that its interaction with matter
breaks explicitly some symmetry that only massless particles
exhibit, e.g. the gauge freedom for vector fields or the chirality
for fermions.}
\end{itemize}

There are only two fashionable candidates that fulfill the
first condition:

\begin{itemize}
\item{A scalar field $ \varphi;$}
\item{The gravitational field.}
\end{itemize}

The Higgs boson $ \varphi$ was postulated to couple universally with
all kinds of matter. However, still to this day there is no evidence
of its universality, put aside its own existence\footnote{See \cite{Evans}
for alternative exotic scenarios suggested to take the place of Higgs
mechanism in case discovery of Higgs boson fails.}. The other one,
gravity, is known to couple with all forms of matter and energy; its
universality is recognized as a scientific truth. We note that after
accepting either one of these two fields as a good candidate that
fulfills the first requirement, it is not a hard job to elaborate
scenarios such that the other two conditions are satisfied too.

In this work we will limit our analysis to the gravitational process \cite{novelloCQG}, \cite{novello}
once in the realm of high-energy physics, the Higgs model produced a
well-known alternative scenario for generating mass for all massive
particles except the Higgs boson itself \cite{self}. Let us just point out a
remarkable property of the Higgs mechanism that within its scenario
was not sufficiently emphasized. It concerns the property that the self-interacting scalar field
in order to generate the mass of the particles must be in its fundamental
state. Its energy distribution is described as
$$ T_{\mu\nu} = V(\varphi_{0}) \, g_{\mu\nu}$$
defining a cosmological constant  $ V(\varphi_{0}).$ However, in this
mechanism, this fact is no further analyzed, since at the realm of
microphysics gravity is ignored. So much for this structure. Let us
turn now to the new mechanism.

\section{Numerical results}

Before entering in the details of the gravitational mechanism let us point out
some of its observational consequences. We start by recalling that the
inverse Compton length of any particle is given in terms of its mass $ M $, the Planck constant $ \hbar $ and light velocity $ c $
yielding
$$ \mu = \frac{c}{\hbar} \, M. $$
For latter use we re-write it in an equivalent way in terms
of gravitational variables using the Newton constant $ G_{N}$ or
equivalently the Einstein constant $ \kappa = 8\pi\, G_{N} / c^{4}.
$ The Schwarzschild solution of the gravitational field of a static
compact object has an horizon -- that is a one-way membrane --
characterized by its Schwarzschild radius
$$ r_{s} = \frac{1}{4\pi} \, \kappa \, M \, c^{2}$$
Using the definition of the Planck length
$$ L_{Pl}^{2} \equiv \frac{1}{8 \pi} \, \kappa \, \hbar \, c $$
it follows that inverse Compton length may be written under the
equivalent form as the ratio between the corresponding Schwarzschild radius and
the square of Planck length:
\begin{equation}
 \mu = \frac{1}{2} \, \frac{r_{s}}{L_{Pl}^{2}}.
 \label{28jun2}
 \end{equation}

The formula of the mass we obtained in \cite{novelloCQG} (and which we will review in the
next section) from the non-minimal coupling of a spinor field $ \Psi$  with
gravity is expressed in terms of the cosmological constant $ \Lambda
,$ the Planck length and parameter $ \sigma $ of the non-minimal
coupling yielding the expression
\begin{equation}
\mu =  \frac{1}{8 \pi} \, \frac{\sigma \, \Lambda}{ L_{Pl}^{2}} \label{28jun1}
\end{equation}
This expression relates two parameters: the mass $ M $ and the
associated non-minimal coupling constant with gravity  $ \sigma$
that has the dimensionality of a volume. The knowledge of one of
these two parameters ( $ M $ and $ \sigma$ ) allows the knowledge of
its companion. By comparison of the above two expressions of  $ \mu,
$ that is, Compton definition eq. (\ref{28jun2}) and our formula for
the mass eq. (\ref{28jun1}) yields the expression of $ \sigma: $

\begin{equation}
\sigma = 4 \, \pi \, \frac{r_{s}}{\Lambda} \label{28jun3}
\end{equation}

Thus different fermion particles that have different masses have different
values of $ \sigma.$ We note furthermore that the ratio $ M / \sigma
$ which has the meaning of a density of mass is a universal constant
given only in terms of $\kappa$ and $ \Lambda.$ How to interpret
such universality? There is a direct and simple way that is the
following. We re-write this formula as a density of energy, that is

\begin{equation}
\frac{M \, c^{2}}{\sigma} = \frac{\Lambda}{\kappa}
\label{16nov}
\end{equation}

The right-hand side is nothing but the density of energy of the
vacuum. Thus we can say that  $ \sigma $ is the volume in which an
homogeneous distribution of the particle energy spreads having the
same value of the vacuum energy density provided by the cosmological
constant, that is, $ \Lambda / \kappa.$

Once our formula of mass for fermions contains gravitational quantities which are
well-known to be extremely small, let us compare it with actual
numbers that we can get, for instance, from the simplest example of the electron.
The main question is: should the coupling constant $\sigma$
become an enormously big value in order to compensate the weakness
of the gravitational field? A direct calculation for the known
elementary particles show that this is not the case. This is a direct consequence, as we shall see in the next section, of the
fact that, in the process of give mass, gravity enters only as a catalyst.  Indeed, for the
simple stable lepton, the electron we find that
$$ r_{s} \approx 1.35 \times 10^{- 55} cm, $$
which implies\footnote{All the values used here were taken from the \textit{Particle Data Group} \cite{pdg}.} that
$$ \sigma_{e} \approx 125 \, cm^{3}.$$

The substance that we call electron is tremendously concentrated
within its Compton wavelength $ \lambda_{c}. $ Indeed if we compare the
density of energy $ M_{e} \, c^{2} / r^{3} $ for $ r = \lambda_{c} $ and $
\sigma, $ it follows that all of the electron is concentrated in its
Compton interior:

$$ \frac{\varrho_{c}}{\varrho_{\sigma}} \approx 10^{31}. $$

Before ending this section let us make a remark in order to test the coherence of our formula (\ref{16nov}) in the cosmological scenario.
Indeed, suppose the extremal case identifying $\sigma$ with the total volume of the Universe. Assuming that
the universe is roughly made by protons, we can estimate the total number of protons $ N_{p}$ in the universe. A direct
 calculation using equation (\ref{16nov}) for $\sigma\approx(10^{28}cm)^{3}$, yields
\begin{equation}
\nonumber
N_{p}\approx 10^{80}\, \mbox{protons}.
\end{equation}
We note that this number is precisely Eddington number.

\section{Minimal mass value}

The present method of evaluating the mass takes into account only classical gravitational aspects.
Thus, in principle it stops to be applied at the quantum level. Indeed, quantum effects become non negligible at least at the
Compton wavelength of a given particle. This means that there is a threshold of applicability of our mechanism. In other words the value of the
length associated to the gravitational mechanism must be higher than the corresponding Compton wavelength of the particle. This
led naturally to the minimum value of the mass of any fermion - call it $ M_{q}$ - that can be generated by the gravitational procedure. In
other words we must have $\sigma \geq \lambda_{c}^{3}, $ that is
$$ M_{q} c^2 \geq \frac{\Lambda}{\kappa} \, \frac{\hbar^3}{M^3_{q} c^3}, $$
from which we obtain that the minimum possible value for the mass is
$$ M_{q} \geq 2.36 \times 10^{-3} \, eV.$$
Thus there is no possibility of having a fermion which has a mass lower than $M_{q}.$

\section{From Mach principle to the new gravity mechanism}

Although a widespread formulation --- identified as Mach\rq s
principle -- that the mass of a body may depend on the overall
properties of the rest-of-the-universe and consequently to gravity,
the association of this dependence to the smallness of gravitational
phenomena was at the origin of the general attitude of disregarding
any possibility to attribute to gravity an important role in the
generation of mass for all bodies (see however \cite{irina}).

This apparent difficulty is eliminated by two steps:

\begin{itemize}
\item{A direct coupling of matter to the curvature of space-time;}
\item{The existence of a vacuum distribution or cosmological
constant  $\Lambda.$}
\end{itemize}

This idea was developed recently \cite{novelloCQG} thus providing a reliable mechanism
by means of which gravity is presented as truly responsible for the
generation of the mass. As a result of such procedure, the final
expression of mass depends neither on the intensity nor on the
specific properties of the gravitational field. This circumvents all
previous criticism against the major role of gravity in the origin
of mass.

The model uses a slight modification of Mach\rq s principle. Let us
remind  that, following Einstein \cite{Einstein}, we can understand
by this principle the statement according to which the entire
inertia of a massive body is the effect of the presence of all other
masses, deriving from a kind of interaction with the latter or, in
other words, the inertial properties of a body $\mathbb{A }$ are
determined by the energy throughout all space. The simplest way to
implement this idea is to consider the state that takes into account
the whole contribution of the rest-of-the-universe onto $\mathbb{A
}$ as the most homogeneous one. Thus it is natural to relate it to
what Einstein attributed to the cosmological constant or, in modern
language, the vacuum of all remaining bodies. This means to describe
the energy-momentum distribution of all complementary bodies of
$\mathbb{A }$ in the Universe under the form
\begin{equation}
T_{\mu\nu}(U) = \frac{\Lambda}{\kappa} \, g_{\mu\nu} \label{17abril}
\end{equation}
Note that this distribution of the energy content of the environment
of the body $\mathbb{A }$ is similar to the Higgs case, although
there is an important distinction concerning the role of this
homogeneous distribution of energy on the  generation of mass: as we
pointed out above, Higgs\rq proposal does not go further to explore
the consequences of this distribution of energy, since it is not
followed by the analysis relating such energy to gravitational
processes. We consider the very fundamental framework dealing with
the basic constituents of matter, the true building blocks, and
treat matter generically as representations of the Lorentz group. In
the present paper we limit our description to the case in which body
$\mathbb{A }$ is identified with fermions.

\subsection{The case of fermions}

 The massless
 theory for a spinor field is given by Dirac equation:
\begin{equation} i\gamma^{\mu} \partial_{\mu} \, \Psi  = 0 \label{221}
\end{equation}
This equation is invariant under $\gamma^{5} $ transformation. In
order to have mass for the fermion this symmetry must be broken. Who
is the responsible for this? Electrodynamics appears in gauge theory
as a mechanism that preserves a symmetry when one pass from a global
transformation to a local one (space-time dependent map). Nothing
similar with gravity. Following \cite{novelloCQG} we assume the idea
that gravity is the true responsible to break the symmetry.

In the framework of General Relativity the non-minimal gravitational
interaction of the fermion is driven by the Lagrangian
$$ L =  L_{D} + L_{int} + L_{\Lambda} + L_{ct}$$
that is
\begin{eqnarray}
L &=& \frac{i}{2} \bar{\Psi} \gamma^{\mu} \nabla_{\mu} \Psi -
\frac{i}{2} \nabla_{\mu} \bar{\Psi} \gamma^{\mu} \Psi \nonumber \\
&+& \frac{1}{\kappa} \,  (1 + \frac{\sigma}{4} \, \Phi)^{-2} \, R -
\frac{1}{\kappa}
 \, \Lambda \nonumber
\\ &-&    \, \frac{3}{8 \kappa} \, \sigma^{2} \,  (1 + \frac{\sigma}{4}  \, \Phi)^{-4} \, \partial_{\mu} \Phi \, \partial^{\mu} \Phi,
\label{29junho3}
\end{eqnarray}
where the non-minimal coupling of the spinor field with gravity is
contained in the term $ V(\Phi) = 1 + \sigma \, \Phi/4 $ that depends
on the scalar $$ \Phi \equiv \bar{\Psi} \, \Psi, $$
which preserves the gauge invariance of the theory under the map $
\Psi \rightarrow \exp(i \, \theta) \, \Psi.$  Note that the
dependence on $ \Phi$ on the dynamics of $ \Psi$ breaks the chiral
invariance of the mass-less fermion, a condition that is necessary
for a mass to appear. The constant $
\sigma$ which has dimensionality $(length)^{3}$ given by
(\ref{28jun3}) is the responsible for the non-minimal coupling and
the presence of the self-interacting term.

This dynamics represents a massless spinor field coupled
non-minimally with gravity. The cosmological constant represents the
influence of the rest-of-the-universe on $\Psi.$

Independent variation of $\Psi$ and $g_{\mu\nu}$ yields
\begin{equation}
 i\gamma^{\mu} \nabla_{\mu} \, \Psi  +  \Sigma \, \Psi = 0, \label{223}
\end{equation}

\begin{equation}
\alpha_{0} \, ( R_{\mu\nu} - \frac{1}{2} \, R \, g_{\mu\nu} ) = -
T_{\mu\nu},
 \label{224}
\end{equation}
where  $ \Sigma$  depends on the curvature scalar $ R $ and on $
\Phi.$

The energy-momentum tensor is defined by

 $$T_{\mu\nu} = \frac{2}{\sqrt{- g}} \, \frac{\delta ( \sqrt{-g} \,
 L)}{\delta g^{\mu\nu}}. $$

Taking the trace of equation (\ref{224}) and inserting it on the expression $\Sigma$ one obtains after some
algebraic manipulation\footnote{See \cite{novelloCQG} for details.} that the equation for the spinor becomes
\begin{equation} i\gamma^{\mu} \nabla_{\mu} \, \Psi  - M \Psi= 0, \label{15}
\end{equation}
where \begin{equation}
 M = \frac{\sigma \, \Lambda}{\kappa \, c^{2}}.
 \label{30julho13}
 \end{equation}

Thus as a result of the above process  the spinor field acquires a
mass $ M $ that depends crucially on the existence of $ \Lambda.$ If
$ \Lambda $ vanishes then the mass of the field vanishes. The
 non-minimal coupling of gravity with the spinor field corresponds to
a specific self-interaction. The mass of the field appears only if
we take into account the existence of all remaining bodies in the
universe --- represented by the cosmological constant. The values of
different masses for different fields are contemplated in the
parameter $ \sigma.$

 This procedure allows us to state that the mechanism proposed here is to be
understood as a realization of Mach principle according to which the
inertia of a body depends on the background of the
rest-of-the-universe. This strategy can be applied in a more general
context in support of the idea that (local) properties of
microphysics may depend on the (global) properties of the universe. In the case $ \sigma = 0 $ the Lagrangian reduces to a massless
fermion satisfying Dirac\rq s dynamics plus the gravitational field
described by General Relativity.

\subsection{A new interpretation of the non-minimal coupling}

 There is another interpretation of
the Lagrangian (\ref{29junho3}) that is worth to point out here because it shows that the non-minimal
coupling described by (\ref{29junho3}) can be interpreted as the conformal coupling of a
scalar field with gravity. Let
us define the non dimensional scalar field $ X $ by setting
$$ X = \frac{1}{\sqrt{6}} (1 + \frac{\sigma}{4} \, \Phi)^{-1} $$
Then, in terms of this new quantity the dynamics (\ref{29junho3}) can be re-written
as
\begin{equation}
L = L_{D} - \frac{\Lambda}{\kappa} - \frac{1}{\kappa} \, (
\partial_{\mu} X \, \partial^{\mu} X - \frac{1}{6} \, R \, X^{2}
), \label{29junho1}
\end{equation}
which is nothing but Dirac dynamics plus the equation of a scalar field $ X $ coupled
in a conformal way to the curvature of space-time. We recognize here the standard procedure of
conformal coupling a scalar field with gravity. When the field $ X $ is identified with the chiral
dependent term constructed with the spinor field through the above definition then the net effect of gravity through the existence of a cosmological constant appears and provides mass  for $ \Psi.$

\section{Modified Mach Principle}

The various steps of our mechanism driven by Lagrangian (\ref{29junho3})
are synthesized as follows:
\begin{itemize}
\item{The dynamics of a massless spinor field $ \Psi$ interacting to gravity in a conformal way is contained in
the Lagrangian
$$ L = L_{D} - \frac{\Lambda}{\kappa} - \frac{1}{\kappa} \, (
\partial_{\mu} X \, \partial^{\mu} X - \frac{1}{6} \, R \, X^{2}
) ; $$}
\item{Gravity is described by General Relativity;}
\item{The action of the rest-of-the-universe on the spinor field,
 through the gravitational intermediary, is contained in the form
of an additional constant term on the Lagrangian noted as $ \Lambda
;$ }
\item{As a result of this process, the field $ \Psi$  acquires a mass
$ M $ given by expression (\ref{30julho13}) and is zero only if the
cosmological constant vanishes;}

\item{This process is completely independent from the intensity and the specific
configuration of the gravitational field.}
\end{itemize}

The generalization of this procedure for all other kinds of matter
which are representations of the Lorentz group (scalar or
tensor fields) can be made straightforwardly through the same lines
as in the present case.

The mechanism presented in this paper allows us to interpret the
mass of any body as being nothing but a local property originated by
the influence of the whole universe intermediated by the
gravitational interaction. In other words, to explain the origin of
mass for all bodies, there is no need to introduce extra fields as,
for instance, the Higgs boson.

 \vspace{0.50cm}
\textbf{Acknowledgements}
 \vspace{0.50cm}

 MN would like to thank FINEP, CNPq and Faperj for financial support and EB thanks CNPq.
We would like also to thank J. M. Salim for many enthusiastic conversations on the subject
of this paper. We acknowledge the staff of ICRANet at Pescara were this paper was done.


\begin{thebibliography}{50}
\bibitem{self}
A curious difficulty of principle appears here: Higgs mechanism
intends to provide mass for all existing particles. However, once a
mass-term appears in the Higgs potential, one should be concerned
with the question: and what is the origin of the Higgs mass? This
problem is completely absent in the case of gravity, which is a long
range field.
\end{thebibliography}
\end{document}